\documentclass[sigconf,natbib=true,anonymous=false]{acmart}
\AtBeginDocument{%
  }

\setcopyright{acmlicensed}
\copyrightyear{2026}
\acmYear{2026}
\setcopyright{cc}
\setcctype{by}
\acmConference[RecSys '26]{20th ACM Conference on Recommender Systems}{September 27-October 02, 2026}{Minneapolis, MN, USA}
\acmBooktitle{20th ACM Conference on Recommender Systems (RecSys '26), September 27-October 02, 2026, Minneapolis, MN, USA}
\acmDOI{10.1145/3773078.3831801}
\acmISBN{979-8-4007-2284-4/2026/09}

\usepackage{multirow}
\usepackage{booktabs}
\usepackage{caption}
\usepackage{enumitem}
\usepackage{algorithm}
\usepackage{algpseudocode}
\usepackage{array}

\begin{document}

\title{Position Bias Undermines Preference Consistency in Listwise LLM-Based Reranking}

\author{Ethan Bito}
\affiliation{%
 \institution{RMIT University}
 \city{Melbourne}
 \country{Australia}}
\email{s4102812@student.rmit.edu.au}

\author{Yongli Ren}
\affiliation{%
 \institution{RMIT University}
 \city{Melbourne}
 \country{Australia}}
\email{yongli.ren@rmit.edu.au}

\author{Estrid He}
\affiliation{%
 \institution{RMIT University}
 \city{Melbourne}
 \country{Australia}}
\email{estrid.he@rmit.edu.au}

\renewcommand{\shortauthors}{Ethan Bito, Yongli Ren, and Estrid He}


\begin{abstract}
Large language models (LLMs) have emerged as promising listwise rerankers for recommender systems, but their reliability under equivalent candidate permutations remains unclear. Since recommendation candidates form an unordered set, a reranker should not depend on the arbitrary order used to serialize them. However, decoder-only LLM rerankers can allow input order to affect model scores, pairwise preferences, and rankings. We study how position bias affects the ranking process induced by LLM-based rerankers. Instead of measuring only changes in final ranked lists, we treat rankings produced under equivalent candidate permutations as observations of an induced preference system. We introduce an evaluation framework measuring pairwise preference instability, global preference inconsistency, and listwise output consistency. This framework characterizes candidate-order sensitivity at the pairwise, global, and output levels. Experiments across multiple LLMs, datasets, and list lengths show that these consistency measures are closely aligned, but can diverge from recommendation effectiveness and marginal position-exposure bias. Improving relevance or flattening exposure across positions does not necessarily restore stable pairwise preferences, globally coherent preference structures, or consistent ranked outputs. These results show that reducing marginal exposure skew is insufficient to establish ranking-function validity in LLM-based reranking. Code is available at \url{https://github.com/ejbito/InvariRank}.
\end{abstract}



\maketitle

\section{Introduction}
\label{sec:introduction}

Large language models (LLMs) have shown strong potential in recommendation, acting as listwise rerankers that jointly score candidate items conditioned on user context \cite{hou2024largelanguagemodelszeroshot,ma2023largelanguagemodelsstable,sun2024chatgptgoodsearchinvestigating,qin2024largelanguagemodelseffective}. In a typical two-stage pipeline, an efficient retriever first generates a candidate set, and the LLM reranks these candidates using user history and item metadata \cite{geng2022recommendation,wei2024llmrec,Dai_2023,sanner2023largelanguagemodelscompetitive,liu2023chatgptgoodrecommenderpreliminary,yue2023llamarectwostagerecommendationusing}.

Despite this promise, LLM-based rerankers exhibit a critical reliability issue. The predicted ranking can change when the order of the candidate set is permuted, even though the underlying set of items remains identical \cite{hou2024largelanguagemodelszeroshot,tang2024found,xu2024incontextexampleorderingguided}. This violates a fundamental assumption in recommendation, where candidates form an unordered set and rankings should not depend on arbitrary input serialization.

Prior work has studied this phenomenon as position bias, and proposed mitigation strategies based on permutation aggregation, calibration, training objectives, and sequential selection \cite{hou2024largelanguagemodelszeroshot,ma2023largelanguagemodelsstable,tang2024found,ren2024selfcalibratedlistwisererankinglarge,chao-etal-2024-make,bito2025evaluatingpositionbiaslarge}. These approaches can reduce order sensitivity, but evaluation often focuses on recommendation effectiveness, marginal exposure across input positions, or agreement between final ranked lists rather than how candidate-order sensitivity appears in the induced preference structure.

In this work, we study position bias as a failure of the induced ranking process. Each ranked list implies pairwise preferences, and preferences aggregated across permutations may be difficult to reconcile with a single global ordering. We distinguish candidate-order dependence from marginal position-exposure skew, which measures how input positions affect top-ranked placement. This raises three research questions: (i) whether local pairwise preferences remain stable across candidate permutations, (ii) whether the induced preference system remains globally coherent, and (iii) how preference-level consistency relates to ranked-output consistency, recommendation effectiveness, and marginal position-exposure bias.

To address these questions, we treat rankings produced under equivalent candidate permutations as observations of an induced preference system and introduce a multi-level framework connecting candidate-order sensitivity to ranking-function validity.

Our experiments across multiple models, datasets, and list lengths show that these three consistency measures give closely aligned assessments of order sensitivity. In contrast, recommendation effectiveness and marginal position-exposure bias can diverge from them: methods that improve relevance or flatten exposure across positions may still produce unstable preferences and ranked outputs. These findings show that marginal exposure correction alone is insufficient to establish permutation-consistent LLM reranking.
\section{Related Work}

Prior to the emergence of prompt-based LLM recommenders, neural recommendation methods incorporated contextual item representations and time-aware self-attention to model user preferences and sequential behaviour~\cite{he2019joint,he2020timesan}. LLMs have been explored for ranking and recommendation, where user histories and candidate items are represented through natural language prompts. Prior work includes both tuning-based and prompting-based approaches, with tuning methods adapting LLMs using task-specific supervision and prompting methods relying on zero- or few-shot inference~\cite{Bao_2023,friedman2023leveraginglargelanguagemodels,wu2024surveylargelanguagemodels,yang2023palrpersonalizationawarellms}. Several studies show LLMs can perform competitively in recommendation and ranking, especially as rerankers or near cold-start recommenders~\cite{Dai_2023,liu2023chatgptgoodrecommenderpreliminary,sanner2023largelanguagemodelscompetitive}.

LLM behavior is also known to be sensitive to input structure. Prompt formatting and the ordering of in-context examples can affect model outputs~\cite{xu2024incontextexampleorderingguided}, and related positional effects have been observed in question answering, retrieval, and long-context reasoning~\cite{liu2023lostmiddlelanguagemodels,hsieh2024middlecalibratingpositionalattention}. These findings raise concerns when unordered recommendation candidate sets are serialized into prompts.

More closely related to our setting, recent work has examined instability and position bias in LLM-based recommendation ranking. Existing mitigation strategies include inference-time aggregation~\cite{hou2024largelanguagemodelszeroshot}, sequential selection~\cite{bito2025evaluatingpositionbiaslarge}, calibration-based correction \cite{ma2023largelanguagemodelsstable}, and fine-tuned order-robust rerankers \cite{chao-etal-2024-make,10.1145/3805712.3809952}. Our work is complementary to these efforts. Rather than proposing a mitigation method, we study candidate-order sensitivity as a reliability issue in listwise LLM reranking. We characterize the resulting failure at the local pairwise, global preference, and ranked-output levels, and examine how these related consistency measures compare with recommendation effectiveness and marginal position-exposure bias.
\section{Methodology}

\subsection{Ranking-Function Validity}

We study listwise reranking with a user context $H$ and candidate set
\[
C = \{c_1, c_2, \ldots, c_K\}.
\]
A reranker takes $H$ and a serialized ordering of $C$ as input, then produces a ranking over the candidates. Since $C$ is an unordered set, a coherent ranking function should depend on the user context and candidate contents rather than the arbitrary serialization used to present the candidates.

Let $\pi_1, \ldots, \pi_M$ denote $M$ sampled permutations of $C$, and let $R_t$ be the ranking produced under permutation $\pi_t$. These rankings form the sampled rankings
\[
\mathcal{R}(H,C) = \{R_1, R_2, \ldots, R_M\}.
\]
Each ranking induces pairwise preferences. For a candidate pair $(c_i,c_j)$, we define
\[
q(i,j)
=
\frac{1}{M}
\sum_{t=1}^{M}
\mathbb{I}[c_i \succ_t c_j],
\]
where $c_i \succ_t c_j$ indicates that $c_i$ is ranked above $c_j$ in $R_t$. The induced preference system is
\[
Q = \{q(i,j)\}_{i \neq j}.
\]

We use $Q$ and $\mathcal{R}(H,C)$ to evaluate whether the reranker behaves consistently across equivalent serializations.

\subsection{Pairwise Preference Instability}

Pairwise Preference Instability (PPI) measures whether the relative preference between two candidates changes as their input positions vary across equivalent permutations. Conditioning on every ordered position pair is sparse because there are $K(K-1)$ possible pairs and only $M$ sampled permutations. We therefore group positions into coarse buckets $\mathcal{B}$, such as head, middle, and tail.

Let $b_t(i)$ denote the bucket containing candidate $c_i$ under permutation $\pi_t$. For a bucket pair $(b_i,b_j)$, let
\[
\mathcal{T}_{ij}(b_i,b_j)
=
\{t \mid b_t(i)=b_i,\ b_t(j)=b_j\}.
\]
The bucket-conditioned preference probability is
\[
q(i,j \mid b_i,b_j)
=
\frac{1}{|\mathcal{T}_{ij}(b_i,b_j)|}
\sum_{t \in \mathcal{T}_{ij}(b_i,b_j)}
\mathbb{I}[c_i \succ_t c_j].
\]
The pairwise instability for candidate pair $(c_i,c_j)$ is
\[
\mathrm{PPI}(i,j)
=
\max_{b_i,b_j \in \mathcal{B}}
q(i,j \mid b_i,b_j)
-
\min_{b_i,b_j \in \mathcal{B}}
q(i,j \mid b_i,b_j).
\]
The maximum and minimum are computed over bucket pairs observed at least once. Averaging over all candidate pairs gives
\[
\mathrm{PPI}
=
\frac{1}{\binom{K}{2}}
\sum_{1 \leq i < j \leq K}
\mathrm{PPI}(i,j).
\]
Lower PPI indicates greater stability across input regions.

\subsection{Global Preference Inconsistency}

Global Preference Inconsistency (GPI) measures whether the induced preference system is globally coherent. Pairwise preference stability does not necessarily imply that $Q$ is well explained by a single ranking over the candidate set.

Let $\sigma$ denote a ranking, where $\sigma(i) < \sigma(j)$ means that $c_i$ is ranked above $c_j$. The disagreement between $\sigma$ and $Q$ is
\[
D(\sigma)
=
\sum_{1 \leq i < j \leq K}
\left(
\mathbb{I}[\sigma(i) < \sigma(j)](1-q(i,j))
+
\mathbb{I}[\sigma(j) < \sigma(i)]q(i,j)
\right).
\]
This quantity measures the pairwise preference mass that contradicts $\sigma$. The best-fitting ranking is
\[
\sigma^\ast
=
\arg\min_{\sigma} D(\sigma).
\]
This is a weighted Kemeny rank-aggregation objective, with $q(i,j)$ defining the pairwise disagreement weights \cite{10.1145/371920.372165}.

Since exact minimization requires searching over all rankings, we initialize $\hat{\sigma}$ by sorting candidates in descending aggregate pairwise win score
\[
s_i=\sum_{j\neq i} q(i,j),
\]
breaking ties by candidate identifier. We then repeatedly apply, over all candidate pairs, the pairwise swap that gives the largest reduction in $D(\sigma)$, stopping when no improving swap remains.

We define GPI as
\[
\mathrm{GPI}
=
\frac{1}{\binom{K}{2}}
D(\hat{\sigma}),
\]
where $\hat{\sigma}$ is the approximate best-fitting ranking. Lower GPI indicates that the induced preference system is well explained by a single global ranking.

\begin{figure*}[t]
    \centering
    \caption{Top-5 exposure by input position on MovieLens-32M using Llama-3.2-3B-Instruct for list lengths $K \in \{15,25,50\}$.}
    \includegraphics[width=0.9\linewidth]{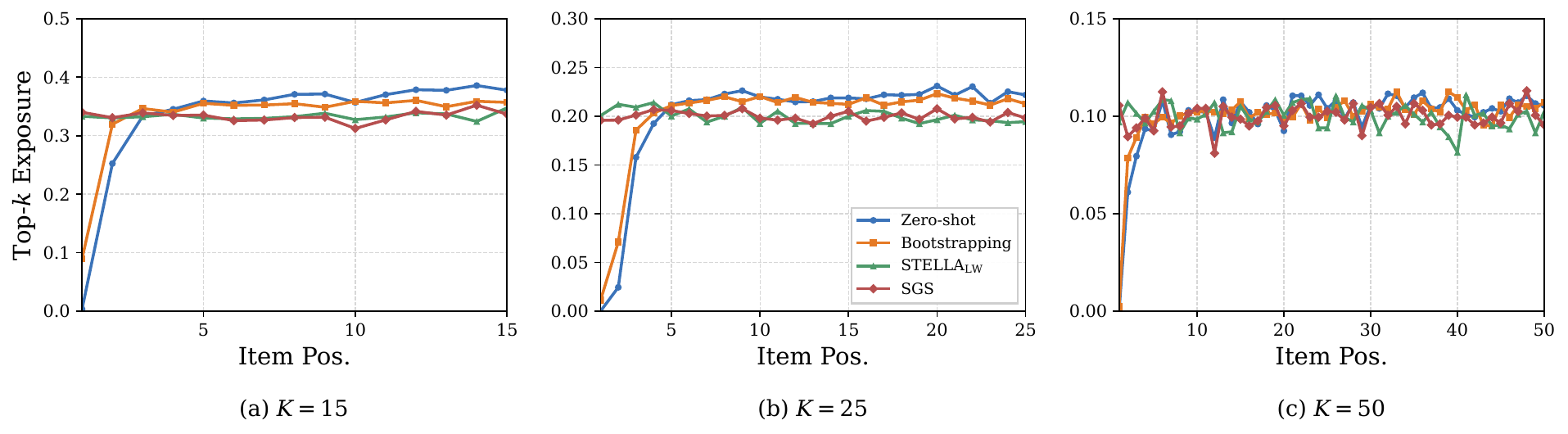}
    \label{fig:position_exposure}
    \vspace{-0.5cm}
\end{figure*}

\subsection{Listwise Output Consistency}

PPI and GPI characterize the preference structure induced across candidate permutations. At the listwise level, we evaluate whether equivalent permutations produce consistent complete rankings.

Given the ranking family $\mathcal{R}(H,C)$, we measure Listwise Output Consistency (LOC) using average pairwise agreement between sampled rankings. We primarily report Kendall's $\tau$. Higher values indicate that the reranker produces more stable ranked outputs across equivalent candidate permutations.

PPI, GPI, and LOC are not intended to be statistically independent. Rather, they provide local, global, and output-level characterizations of the same candidate-order sensitivity. Unlike PPI and GPI, which evaluate the induced preference system, LOC measures consistency directly at the level of observable ranked outputs.

\subsection{Marginal Position-Exposure Bias}

We additionally measure how input position affects placement near the top of the ranked list. For permutation $\pi_t$, let $p_t(i)$ denote the input position of candidate $c_i$. We define top-$k$ exposure at input position $p$ as
\[
E_k(p)
=
\Pr\left(c_i \in \operatorname{Top}\text{-}k(R_t)
\mid p_t(i)=p\right).
\]
This probability is estimated across candidates, queries, and sampled permutations. Since candidate positions are randomized, a position-insensitive reranker should exhibit approximately uniform exposure, with $E_k(p)=k/K$ for each input position. Systematic deviations from this baseline indicate marginal position-exposure bias. Unlike PPI, GPI, and LOC, this measure captures average top-ranked placement by input position rather than consistency across equivalent permutations.
\section{Experimental Setup}

\textbf{Datasets.}
We evaluate on MovieLens-32M and Amazon Books. User interactions are ordered chronologically and split temporally into training and evaluation sets. For each evaluation query, the user history is constructed from past interactions, and held-out interactions are used as relevant items. Each evaluated candidate set contains at least one held-out relevant item.

\textbf{Retrieval.}
Candidate sets are constructed using LightGCN as a first-stage retriever~\cite{he2020lightgcnsimplifyingpoweringgraph}. We follow a standard two-stage recommendation pipeline~\cite{yue2023llamarectwostagerecommendationusing,gao2025llm4rerankllmbasedautorerankingframework}, in which a first-stage recommender produces a candidate set and an LLM reranks it. For each query, we take the top-$K$ candidate items for LLM reranking, with $K \in \{15,25,50\}$.

\textbf{Models and ranking extraction.}
We evaluate three instruction-tuned decoder-only LLMs: Llama-3.2-3B-Instruct, Mistral-7B-Instruct-v0.3, and Qwen2.5-7B-Instruct. All models use the same prompt template. Evaluation rankings are extracted from token log-probabilities over candidate marker identifiers instead of sampled text generation~\cite{10.1145/3805712.3809952,zhong-chen-2021-frustratingly}. This makes ranking deterministic for a fixed serialized input and avoids failures from parsing generated ranked lists. Consequently, variation across evaluation rankings is induced by candidate permutation rather than decoding randomness or formatting errors. For each query and method, we evaluate $M=20$ candidate permutations.

\textbf{Baselines.}
We compare zero-shot listwise reranking with three position-bias mitigation baselines. Bootstrapping aggregates rankings from independently shuffled candidate orders using Borda count~\cite{hou2024largelanguagemodelszeroshot}. SGS iteratively selects the highest-ranked candidate, removes it, and reshuffles the remaining candidates~\cite{bito2025evaluatingpositionbiaslarge}. STELLA$_{\mathrm{LW}}$ adapts STELLA~\cite{ma2023largelanguagemodelsstable} to listwise ranking. Its generation-based calibration stage uses up to 150 probing candidate sets, places each relevant candidate at every input position, and repeats each placement five times with shuffled non-target candidates to estimate a smoothed position-transition matrix. At inference, STELLA$_{\mathrm{LW}}$ maintains a candidate-level Bayesian posterior and updates it using the input position of the model's top-ranked candidate over the original and shuffled candidate orders. It performs up to ten updates with early stopping, returns the ranking induced by the minimum-entropy posterior, and falls back to the raw ranking when the posterior information gain is negligible. Remaining settings are provided in the repository.
\begin{table*}[t]
\centering
\caption{Reranking effectiveness and consistency on MovieLens-32M and Amazon Books with $K=25$. Higher is better for HR@5, nDCG@5, and LOC (Kendall's $\tau$); lower is better for PPI and GPI.}
\label{tab:main_results}
\small
\setlength{\tabcolsep}{4pt}
\renewcommand{\arraystretch}{0.9}

\vspace{-0.2cm}
\begin{tabular}{clccccc@{\hspace{5pt}}ccccc}
\toprule
& &
\multicolumn{5}{c}{\textbf{MovieLens-32M}} &
\multicolumn{5}{c}{\textbf{Amazon Books}} \\
\cmidrule{3-7}
\cmidrule{8-12}
\multicolumn{1}{c}{\textbf{Model}}
& \multicolumn{1}{c}{\textbf{Method}}
& \textbf{HR@5} $\uparrow$
& \textbf{nDCG@5} $\uparrow$
& \textbf{PPI} $\downarrow$
& \textbf{GPI} $\downarrow$
& \textbf{LOC} $\uparrow$
& \textbf{HR@5} $\uparrow$
& \textbf{nDCG@5} $\uparrow$
& \textbf{PPI} $\downarrow$
& \textbf{GPI} $\downarrow$
& \textbf{LOC} $\uparrow$ \\
\midrule






\multirow[c]{4}{*}{Llama-3B}
& Zero-shot
& 0.5560 & 0.2301 & 0.4775 & 0.1149 & 0.6358
& 0.2145 & 0.1477 & 0.3853 & 0.0983 & 0.6986 \\
& Bootstrapping
& 0.5480 & 0.2256 & \underline{0.2992} & \underline{0.0827} & \underline{0.7475}
& \underline{0.2302} & \textbf{0.1590} & \underline{0.2387} & \underline{0.0668} & \underline{0.7981} \\
& STELLA$_{\mathrm{LW}}$
& \textbf{0.5922} & \underline{0.2318} & 0.8331 & 0.3648 & 0.1181
& \textbf{0.2345} & \underline{0.1583} & 0.8337 & 0.3644 & 0.1201 \\
& SGS
& \underline{0.5634} & \textbf{0.2324} & \textbf{0.1972} & \textbf{0.0586} & \textbf{0.8268}
& 0.2117 & 0.1466 & \textbf{0.1876} & \textbf{0.0571} & \textbf{0.8332} \\

\cmidrule(lr){2-12}

\multirow[c]{4}{*}{Mistral-7B}
& Zero-shot
& 0.4794 & \underline{0.1778} & 0.4726 & 0.1195 & 0.6290
& 0.2325 & 0.1604 & 0.4329 & 0.1147 & 0.6367 \\
& Bootstrapping
& 0.4790 & 0.1771 & \underline{0.2970} & \underline{0.0831} & \underline{0.7481}
& 0.2320 & \underline{0.1613} & \underline{0.2787} & \underline{0.0796} & \underline{0.7523} \\
& STELLA$_{\mathrm{LW}}$
& \textbf{0.5490} & \textbf{0.1917} & 0.8902 & 0.3861 & 0.0616
& \textbf{0.2435} & \textbf{0.1643} & 0.8452 & 0.3410 & 0.1087 \\
& SGS
& \underline{0.4840} & 0.1753 & \textbf{0.2150} & \textbf{0.0643} & \textbf{0.8107}
& \underline{0.2335} & 0.1554 & \textbf{0.1956} & \textbf{0.0591} & \textbf{0.8205} \\

\cmidrule(lr){2-12}

\multirow[c]{4}{*}{Qwen-7B}
& Zero-shot
& 0.5344 & 0.2136 & 0.5408 & 0.1372 & 0.5799
& 0.2530 & 0.1627 & 0.4496 & 0.1219 & 0.6189 \\
& Bootstrapping
& \underline{0.5450} & \textbf{0.2166} & \underline{0.3288} & \underline{0.0938} & \underline{0.7198}
& \underline{0.2535} & \textbf{0.1759} & \underline{0.2844} & \underline{0.0831} & \underline{0.7444} \\
& STELLA$_{\mathrm{LW}}$
& \textbf{0.5528} & 0.2086 & 0.8328 & 0.3615 & 0.1230
& \textbf{0.2594} & \underline{0.1652} & 0.8291 & 0.3342 & 0.1245 \\
& SGS
& 0.5366 & \underline{0.2152} & \textbf{0.2654} & \textbf{0.0804} & \textbf{0.7641}
& 0.2460 & 0.1641 & \textbf{0.2202} & \textbf{0.0661} & \textbf{0.7989} \\

\bottomrule
\end{tabular}
\end{table*}

\section{Results}

\textbf{Effectiveness.}
Table~\ref{tab:main_results} reports effectiveness and consistency at $K=25$. STELLA$_{\mathrm{LW}}$ obtains the highest HR@5 for all three models on both datasets. The best nDCG@5 varies across methods: SGS performs best for Llama-3B on MovieLens-32M, STELLA$_{\mathrm{LW}}$ for Mistral-7B on both datasets, and bootstrapping in the remaining three settings. Thus, even standard effectiveness metrics do not consistently favour the same mitigation method.

\begin{figure}[h]
    \vspace{-0.35cm}
    \centering
    \caption{PPI and GPI across candidate-list lengths for Llama-3.2-3B-Instruct on MovieLens-32M. Lower values indicate greater preference consistency.}
    \includegraphics[width=1\linewidth]{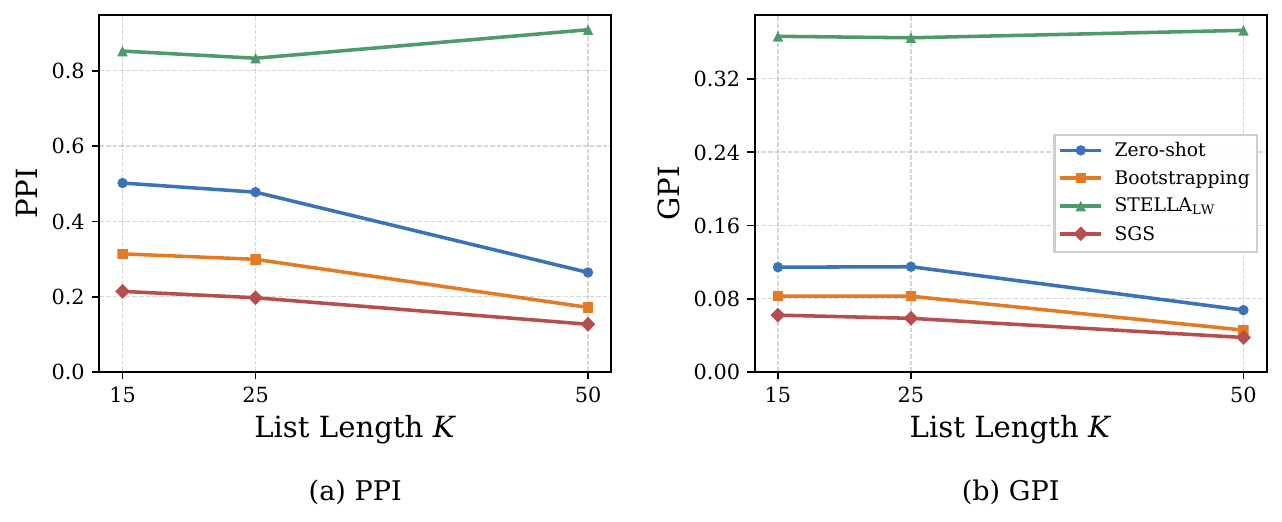}
    \label{fig:ppi_gpi_by_k}
    \vspace{-1.05cm}
\end{figure}

\textbf{Preference consistency.}
PPI and GPI exhibit the same method ordering across both datasets and all three models. Relative to zero-shot reranking, bootstrapping reduces both pairwise instability and global inconsistency, while SGS produces the lowest PPI and GPI throughout. In contrast, STELLA$_{\mathrm{LW}}$ produces the highest PPI and GPI in every setting, despite obtaining the highest HR@5. Figure~\ref{fig:ppi_gpi_by_k} shows that this ordering also persists across $K \in \{15,25,50\}$ for Llama-3B on MovieLens-32M. Improving top-ranked relevance therefore does not necessarily yield stable pairwise preferences or a globally coherent induced preference system.

\textbf{Output consistency.}
LOC follows the same method ordering as PPI and GPI across all six model--dataset combinations. Bootstrapping improves agreement over zero-shot reranking, SGS produces the most consistent ranked outputs, and STELLA$_{\mathrm{LW}}$ produces the lowest LOC, with all values below $0.13$. The agreement among PPI, GPI, and LOC shows that candidate-order sensitivity is reflected consistently at the pairwise, global, and ranked-output levels. Their divergence from HR@5 demonstrates that recommendation effectiveness alone does not establish permutation consistency.

\textbf{Marginal position exposure.}
Figure~\ref{fig:position_exposure} shows top-5 exposure across input positions on MovieLens-32M using Llama-3B for $K \in \{15,25,50\}$. Zero-shot reranking exhibits substantial positional skew, with early input positions receiving lower exposure than later positions. Bootstrapping reduces this skew but does not eliminate it, while STELLA$_{\mathrm{LW}}$ and SGS produce flatter exposure curves.

These results distinguish marginal exposure correction from permutation consistency. STELLA$_{\mathrm{LW}}$ flattens exposure across input positions, yet produces the highest PPI and GPI and the lowest LOC. Reducing average position-dependent exposure is therefore insufficient to recover stable pairwise preferences, a coherent global preference structure, or consistent ranked outputs. SGS improves both marginal exposure and all three consistency measures in the examined setting, showing that these properties can improve together but are not equivalent.

\textbf{Efficiency.}
The methods differ substantially in inference cost. Zero-shot reranking requires one forward pass per candidate set, while bootstrapping aggregates three shuffled passes. With selection size one, SGS requires $K$ sequential passes, corresponding to 25 when $K=25$. STELLA$_{\mathrm{LW}}$ performs up to ten inference-time ranking queries with early stopping and additionally requires a one-off calibration stage. We use up to 150 probing candidate sets in this stage, placing each relevant candidate at every input position and repeating each placement five times.

\textbf{Summary.}
Across both datasets, PPI, GPI, and LOC provide closely aligned assessments of candidate-order sensitivity, but can diverge from recommendation effectiveness and marginal position-exposure bias. STELLA$_{\mathrm{LW}}$ obtains the highest HR@5 while performing worst on all three consistency measures, and produces flatter exposure in the examined MovieLens-32M setting. SGS gives the strongest consistency and comparatively flat exposure, but requires sequential inference. These trade-offs motivate evaluating LLM-based rerankers using effectiveness, permutation consistency, marginal exposure, and inference cost together.

\section{Conclusion}

We study the reliability of listwise LLM-based rerankers under equivalent candidate permutations. By treating the resulting rankings as observations of an induced preference system, we characterize candidate-order sensitivity at the pairwise, global, and ranked-output levels. Across models and datasets, PPI, GPI, and LOC provide closely aligned assessments of permutation consistency, but can diverge from recommendation effectiveness and marginal position-exposure bias. In particular, improving top-ranked relevance or flattening exposure across positions does not necessarily restore stable preferences or consistent rankings. These findings show that marginal exposure correction alone is insufficient to establish ranking-function validity and motivate evaluating LLM-based rerankers using effectiveness and consistency together.

\clearpage

\clearpage

\bibliographystyle{ACM-Reference-Format}
\bibliography{references}


\end{document}